\documentclass{emulateapj}
\usepackage{apjfonts}
\usepackage{epsf}
\begin{document}
\slugcomment{}
\shortauthors{J. M. Miller et al.}
\shorttitle{NGC 4388}

\title{A NICER Look at Strong X-ray Obscuration in the Seyfert-2
  Galaxy NGC 4388}

\author{J.~M.~Miller\altaffilmark{1},
  E. Kammoun\altaffilmark{1},
  R.~M.~Ludlam\altaffilmark{1},
  K.~Gendreau\altaffilmark{2},
  Z.~Arzoumanian\altaffilmark{2},
  E.~Cackett\altaffilmark{3},
  F.~Tombesi\altaffilmark{2.4,5}
}

\altaffiltext{1}{Department of Astronomy, University of Michigan, 1085
  South University Avenue, Ann Arbor, MI 48109-1107, USA,
  jonmm@umich.edu}
\altaffiltext{2}{NASA Goddard Space Flight Center, Code 662,
  Greenbelt, MD, 20771, USA}
\altaffiltext{3}{Department of Physics \& Astronomy, Wayne State
  University, 666 West Hancock Street, Detroit, MI, 48201, USA}
\altaffiltext{4}{Department of Physics, University of Rome ``Tor
  Vergata,'' Via della Ricerca Scientifica 1, I-00133 Rome, Italy}
\altaffiltext{5}{Department of Astronomy, University of Maryland,
  College Park, MD 20742, USA}

\begin{abstract}
We present an analysis of the time-averaged spectrum of the Seyfert-2
active galaxy NGC 4388, obtained by {\it NICER}.  The intrinsic
strength of the reflection spectrum in NGC 4388, the large collecting
area and favorable pass band of {\it NICER}, and a net exposure of
105.6~ks yielded an exceptionally sensitive spectrum.  Using two
independent families of models, the intrinsic spectrum from the
central engine is found to be highly obscured but not Compton-thick.
Enforcing physical self-consistency within each model, the independent
treatments give formally consistent results: $N_{\rm H} =
2.67_{-0.03}^{+0.02}\times 10^{23}~{\rm cm}^{-2}$ or $N_{\rm H} =
2.64^{+0.03}_{-0.03}\times 10^{23}~ {\rm cm}^{-2}$.  Past measurements
made with {\it Suzaku} and {\it XMM-Newton} are in broad agreement
with these column density values.  A more recent measurement with {\it
  NuSTAR} (in late 2013) recorded a column density about twice as large;
the robustness of this variability is reinforced by the use of
consistent models and procedures.  The neutral Fe~K$\alpha$ line in
the {\it NICER} spectrum is nominally resolved and consistent with an
origin in the optical broad line region (BLR).  The data also require
ionized absorption in the Fe K band, similar to the ``warm absorbers''
detected in Seyfert-1 active galactic nuclei (AGN).  The low-energy
spectrum is consistent with a set of ionized plasma components.  We
discuss these findings and note that the geometric inferences that
derive from this analysis can be tested with {\it XRISM} and {\it
  Athena}.
\end{abstract}

\section{Introduction}
The nature of the obscuration that defines Seyfert-2 AGN remains a
matter of study.  The obscuring gas and dust may be associated with
a distant, parsec-scale torus that covers about half of the sky, as
seen from the central engine (e.g., Antonucci 1993).  However, the
obscuration could originate orders of magnitude closer to the black
hole, at radii consistent with the optical broad line region (BLR).
Some models of the BLR suggest that it is also dusty, and that the
enhanced cross section of the dust is responsible for lifting the
material above the plane of the disk (e.g., Czerny \& Hryniewicz 2011,
Czerny et al.\ 2015).

The light crossing time out to a parsec-scale torus is over three
years, and Keplerian time scales at that radius are much longer.
Thus, even variability on a time scale of months is not easy to
reconcile with very distant obscuration.  In a study of obscured AGN,
Risaliti (2002) reported column density variations on the time scale
of months in 23 of 24 cases, typically by factors of 2--3.  NGC 4388
may represent one of the most extreme examples of such variability.
In this AGN, Elvis et al.\ (2004) found significant absorption
variability between observations separated by hours.  That specific
episode was an ``unveiling event'', wherein the obscuring column was
dramatically reduced.  It is reasonable to expect, then, that the
source could also display periods of greatly enhanced obscuration.
However, despite widespread study and strong variability, NGC 4388 has
never been observed in a Compton-thick state ($N_{\rm H}\geq 1.5\times
10^{24}~{\rm cm}^{-2}$).  This may suggest that special conditions,
such as chance alignment of distant clouds, are unlikely to cause a
source to be observed as Compton-thick with any regularity.

X-ray emission line spectroscopy provides another angle on the nature
and location of neutral absorption in obscured AGN.  When cold gas is
illuminated by hard X-rays, a characteristic ``reflection'' spectrum
is produced (e.g., George \& Fabian 1991).  The most prominent part of
a reflection spectrum is an Fe K emisison line.  The strength of this
line relative to the observed continuum can encode the fraction of the
central engine that is covered by the reflector, and any velocity
broadening to the line can be used to constrain the radial location of
the obscuring and reflecting gas (for a recent general treatment of
geometrical inferences, see Ramos Almeida \& Ricci 2017).

In a study of {\it Chandra} gratings spectra of Seyfert-1 galaxies,
Shu et al.\ (2010) found that the narrow Fe K emission line is
consistent with being produced in the optical BLR in approximately
half of the cases examined.  This suggests that the BLR is plausibly
the inner edge of the obscuration that is observed in Seyfert-2
galaxies.  In a separate study of 13 Seyfert-1 AGN, Gandhi et
al.\ (2015) found that the dust sublimation radius likely forms an
outer envelope to the neutral Fe K line production zone.  Recent
focused studies of single sources also conclude that cold reflecting
gas (that would be seen as obscuration in a Seyfert-2) is located
within the optical BLR.  The neutral Fe K line in the Seyfert-1 AGN
NGC 7213 is resolved in a {\it Chandra} gratings spectrum, and clearly
links the line to the optical broad line region (Bianche et
al.\ 2008).  In the Seyfert-1.5 AGN NGC 4151, asymmetry in the Fe K
emission line is observed in the {\it Chandra} grating spectrum
(Miller et al.\ 2018), suggesting an origin within the BLR or
potentially smaller radii.

NGC 4388 is an Sb galaxy in the Virgo Cluster.  Recent work has
estimated its distance at $d=18.0$~Mpc (Sorce et al.\ 2014).
In some obscured AGN, measurements of the black hole mass can be
particularly difficult since broad emission lines are either weak or
unobserved.  However, NGC 4388 has an H$_{2}$O megamaser in its disk;
exploiting this gives an extremely precise black hole mass of $M_{BH}
= 8.4\pm 0.2\times 10^{6}~M_{\odot}$ (Kuo et al.\ 2011).  This
excellent measurement is of major help to efforts to understand the
radial location and geometry of obscuration in this source.

Owing to its bright, neutral Fe~K$\alpha$ line and prior evidence of
dramatic absorption variability, NGC 4388 was targeted for early
observations with {\it NICER} (Gendreau et al.\ 2016).  In Section 2,
the data reduction is described.  Section 3 details the spectral
analysis that was undertaken.  In Section 4, the results are
discussed.  

\section{Observations and Reduction}
NGC 4388 was observed on numerous occasions through the early phase of
the {\it NICER} mission.  The first observation was obtained on
2017-12-03 (MJD 58090); this analysis includes all data accumulated on
and before 2019-03-29 (MJD 58571).  The reduction was accomplished
using HEASOFT version 6.25, and the latest calibration files
consistent with that software release.  The data were processed using
\texttt{NICERDAS}.  Using \texttt{nimaketime}, the raw exposures were
filtered to exclude times close to the South Atlantic Anomaly, times
with high particle background background (via the \texttt{COR\_SAX}
parameter), times of high optical loading (via the
\texttt{FPM\_UNDERONLY\_COUNT}), and to avoid bad sun angles (we
required a minimum of 30 and 40 degrees above the Earth limb and
bright Earth limb, respectively).  After filtering, the net exposure
time is 105.6~ks.  The individual event lists were then combined using
the FTOOL \texttt{nicermergeclean}, and the total spectrum was
extracted using the tool \texttt{XSELECT}.

The ISS orbit takes {\em NICER} through a wide range of geomagnetic
latitudes, each with its own background characteristics.  At high
latitudes, the particle background is dependent on space weather and
the variability of the Sun.  Individual observations thus have
different background levels that must be understood to maximize the
science return.  To calibrate the background, {\em NICER} has (to
date) collected over 2.7 Ms of exposure on background fields used and
characterized by {\it RXTE} (Jahoda 1996) and also a few select
locations near some of the faint millisecond pulsars
that are key to the {\em NICER} core science mission.  For this paper,
the background was estimated using the ``space weather'' method
(Gendreau et al., in prep.) and uses environmental data to parse the
background database.  This tool uses a model of the magnetic cut-off
rigidity the (\texttt{COR\_SAX} parameter) and the space weather index
``KP'' (Kennziffer Planetary; Bartels et al.\ 1939).  KP is derived from
a worldwide network of magnetometers that publishes data every 3 hours
[ https://www.swpc.noaa.gov/products/planetary-k-index ].  KP values
range from $0-9$, where low values indicate calm space weather, while
higher values indicate geomagnetic storms. The space weather
background tool builds a library of background spectra divided amongst
these environmental variables to predict a background spectrum for a
given observation.

Figure 1 plots the background-subtracted {\it NICER} light curve of
NGC 4388, and the evolution of the source hardness.  The hardness
curve is the ratio of counts in the 8--12~keV band to counts in the
0.3--2.0~keV band.  In the higher band, the source is dominant and
relatively free of obscuration.  In the lower band, contributions from
the host galaxy are likely to be important and to give a measure of
flux stability, though absorption is important.  The data show a
concentration of low and negative flux points, and negative spectral
hardness, around $t \simeq 1.2$~Ms.  This episode is due to a brief
period of high background that was not caught by the standard
screening criteria.  This interval was excised and not considered for
spectral analysis.  The light curve shows rare intervals with high
flux, but these are sporadic.  The hardness ratio may be a better
indicator of large changes in the source spectrum and/or obscuring
column, but the errors on the hardness curve are large given the low
count rates.  The centroid values are fairly steady, however, with
only a few points that deviate from the envelope of centroid values.
An apparent spike in the count rate about 3.3~Ms into the monitoring
campaign is not highly significant, and a putative delayed reaction in
the hardness ratio is within the errors.  For these reasons, this
initial analysis of NGC 4388 is confined to the time-averaged spectrum
of the source.

Owing to the fact that the spectrum is comprised of very strong
features, and because the modeling is not concerned with, e.g.,
separating relativistic disk reflection features from the continuum,
we have not ``Crab-corrected'' the spectra (see, e.g., Ludlam et
al.\ 2018, Miller et al.\ 2018).  Prior to spectral fitting, the data
were grouped to require a signal-to-noise ratio of 10, using the FTOOL
\texttt{ftgrouppha}.

\section{Analysis and Results}
The spectra were analyzed using XSPEC version 12.10.1 (Arnaud 1996).
A $\chi^{2}$ statistic was minimized in the fitting procedure.  {\it
  NICER} spectra are nominally valid down to 0.2~keV, but we
restricted our fits to the 0.6--10.0~keV range owing to the
limitations of some models (see below).  Initial fits were made using
the standard default weighting within XSPEC; refinements were made
using ``model'' weighting.  All of the errors in this work reflect the
value of the given parameter at its 1$\sigma$ confidence limits.
Uncertainties on the physical models were obtained after running a
Markov Chain Monte Carlo within XSPEC.  The chains started from the
best-fit model, used the Goodman-Weare algorithm (Goodman \& Weare
2010) with 100 walkers, had a length of $4\times 10^{6}$, and a
burn-in of $6\times 10^{5}$.

\subsection{Phenomenological Models}
Simple models, such as an absorbed power-law, fail to fit the full
spectrum.  However, it can be useful to characterize the strength of
the neutral Fe~K$\alpha$ line and the sensitivity at which it is
detected over a narrow band.  Fitting with a simple Gaussian and
power-law in the 5.5--6.9~keV band to avoid broad continuum trends and
the Fe K edge, we measure a line centroid of ${\rm E} =
6.372^{+0.009}_{-0.003}$~keV (in the frame of NGC 4388).  This is
slightly below the laboratory value, but it is within the uncertainty
in the instrumental energy calibration.


The line is nominally resolved (measurements of its width
statistically exclude zero) with $\sigma = 40^{+10}_{-10}$~eV, or
about three times the nominal separation of the Fe~K$\alpha_{1}$ and
Fe~K$\alpha_{2}$ components (13 eV).  This translates to a broadening
of $v = 1200^{+400}_{-400}~{\rm km}~ {\rm s}^{-1}$.  If the broadening
is Keplerian, it may not be the full velocity but rather a projected
velocity.  If the accretion plane is viewed at $\theta = 60^{\circ}$,
for instance, the velocity should be twice larger, $v =
2300^{+800}_{-800}~{\rm km}~ {\rm s}^{-1}$.  This velocity would
correspond to orbits at $r = 1.7_{-0.8}^{+1.8}\times 10^{4}~
GM/c^{2}$, consistent with the optical BLR in many AGN (see, e.g.,
Peterson et al.\ 2004).

Finally, the flux of the line is measured to be $F =
8.3^{+0.6}_{-0.6}\times 10^{-5}~ {\rm ph} ~{\rm cm}^{-2}~ {\rm
  s}^{-1}$.  This translates to an equivalent width of $W =
73^{+5}_{-5}$~eV.  In Figure 2, for instance, which is based on more
physical modeling (see below), it is clear that the line is only a
fraction of the local continuum flux.  In contrast, the neutral
Fe~K$\alpha$ line in Compton-thick AGN is often several times stronger
than the local continuum (e.g., Kammoun et al.\ 2019; however, also
see Boorman et al.\ 2018).

\subsection{Modeling with Pexmon}
%
%
We next pursued modeling with \texttt{pexmon}.  This model is built on
\texttt{pexrav} (Magdziarz \& Zdziarski 1995), which treats cold, neutral
reflection -- minus lines -- from a slab of infinite column density.
Nandra et al.\ (2007) updated \texttt{pexrav} to include Fe and Ni
K$\alpha$ and K$\beta$ lines, with strengths linked to the reflection
in the manner dictated by atomic physics (see, e.g., George \& Fabian
1991).  If a source is not Compton-thick, and if the inclination is not
extreme, it is likely that the viewing angle to the central engine is
obscured while reflected emission from the far side of the central
engine is not as obscured.  Similarly, diffuse plasma emission from
the larger nuclear region should not be obscured.  However, the direct
emission from the central engine should be obscured, with only a
fraction passing through to the observer.  There is an angle
dependence within \texttt{pexmon}; this is the inclination at which
the {\em reflector} is viewed, not the scattering angle, nor the inner
disk inclination.

Within XSPEC, we built our model as follows:\\

\noindent \texttt{phabs[1]*(zphabs[2]*photoion[3]* cutoffpl[4]
  $+$ zphabs[5]*const[6]*cutoffpl[7] $+$ pexmon[8] $+$ mekal[9] $+$
  mekal[10] $+$ mekal[11]}.\\

\noindent Here, \texttt{phabs[1]} is the line of sight absorption in
the Milky Way (fixed to $N_{\rm H} = 4.0\times 10^{21}~ {\rm cm}^{-2}$
in all fits; this value is not high enough to affect any fits in the
0.6--10.0 keV band).  Then, \texttt{zphabs[2]} is the neutral column
density along the line of sight acting on the continuum emission from
the central engine in the source frame, modeled as a cut-off power-law
via \texttt{cutoffpl[4]}.  In between the central engine and the
neutral obscuration, the flux passes through an ionized absorber,
\texttt{photoion[3]} (described in more detail below).  Some IR
studies of Seyfert-2 AGN find evidence of dust in the polar regions
(e.g., Honig et al.\ 2013), so the model allows for a fraction of the
flux from the central engine to be scattered into our line of sight
(via \texttt{constant[6]*cutoffpl[7]}, where the parameters of
\texttt{cutoffpl[7]} are linked to those of \texttt{cutoffpl[4]} and
$0\leq \texttt{contant[6]} \leq 1$).  (In preliminary tests, we found
that the cut-off was poorly constrained, so $E_{\rm cut} = 500$~keV
was enforced in all fits.)  This scattered emission is allowed to pass
through a different column density than that measured along the line
of sight (via \texttt{zphabs[5]}).  Reflection -- nominally from
distant material on the far side of the central engine -- is modeled
with \texttt{pexmon[8]}.  The power-law index, cut-off energy, and
flux normalization within \texttt{pexmon[8]} were linked to the values
in \texttt{cutoffpl[4]}.  Finally, the low energy spectrum required
three diffuse plasma components, modeled with \texttt{mekal}.  These
collisional plasma models are adopted for simplicity; fits to the {\it
  NICER} spectrum are equally good if one of the \texttt{mekal}
components is replaced with a photoionized plasma model.

The photoionized absorption component is a high-resolution XSTAR table
model (e.g., Kallman \& Bautista 2001; also see Miller et al.\ 2015,
2016), generated assuming a standard $\Gamma = 1.7$ X-ray power-law
and a UV blackbody disk.  The UV/X-ray flux ratio was set to be 10:1,
broadly consistent with bolometric corrections to Seyfert X-ray fluxes
(Vasudevan \& Fabian 2009).  A bolometric lumnosity of $L = 10^{44}~
{\rm erg}~ {\rm s}^{-1}$ was assumed.  Assuming a different luminosity
would not directly affect the ionization parameter, since this
measurement is driven by line flux ratios.  Fits to the data without
this model showed evidence of absorption in the 6.6--6.7~keV range,
potentially consistent with He-like Fe~XXV seen in a number of
Seyfert-1s.  The variable parameters that can be constrained with
XSPEC include the absorption column density ($N_{\rm H}$), ionization
(${\rm log}~\xi$, where $\xi = L/nr^{2}$), and the velocity shift.
The fits did not measure a significant shift so this parameter was
fixed to have zero velocity.  Comparing the $\chi^{2}$ fit statistic
to models excluding the XSTAR component, a simple F-test suggests that
the absorber is significant at the $4\sigma$ level of confidence.  

The parameters measured with this model are listed in Table 1, and the
fit is shown in Figure 2.  The confidence contours resulting from the
MCMC error analysis are shown in Figure 3.  The neutral obscuration is
measured to be Compton-thin, with upper limits that exclude the
Compton-thick regime.  The ``reflection fraction'' is constrained to
be much less than unity, again consistent with the source being
Compton-thin.  Emission that is scattered from the polar region, or
another region out of the line of sight, is very small but nonzero.
An iron abundance of $A_{Fe} = 2.0$ relative to solar was slightly
preferred by the data, and fixed in all fits, but is not required when
\texttt{mytorus} is used (see below).  The model detailed in Table 1
is not formally acceptable in a statistical sense; this is largely
driven by remaining calibration uncertainties in the Si~K and Au
M-shell region near 2--3 keV.

Finally, we note that after removing the neutral obscuring column, our
best-fit \texttt{pexmon} model construction implies an unabsorbed flux
of $F \sim 2.21(7)\times 10^{-10}~ {\rm erg}~ {\rm cm}^{-2}~ {\rm
  s}^{-1}$ in the 0.6--10~keV band.  This corresponds to an
0.6--10.0~keV X-ray luminosity of $L = 8.6(3)\times 10^{42}~ {\rm
  erg}~ {\rm s}^{-1}$, or $L_{X}/L_{\rm Edd} = 8.1(3)\times 10^{-3}$.
Bolometric corrections for Seyferts can range between 15--70
(Vasudevan \& Fabian 2007), but they tend to be higher at higher
Eddington fractions.  For the lowest correction factor, an
Eddington fraction of $L_{\rm bol}/L_{\rm Edd.} \simeq 0.12$ is
implied; for the largest factor, $L_{\rm bol}/L_{\rm Edd.} \simeq 0.57$.

\subsection{Modeling with Mytorus}
We also made fits with the \texttt{mytorus} suite (Murphy \& Yaqoob
2009, Yaqoob et al.\ 2010), to add even more physical self-consistency
than is possible with \texttt{pexmon}.  The model is only valid above
0.6~keV.  \texttt{Mytorus} does not include a cut-off energy in
its power-law functions; we used versions of the obscuring, scattered,
and line files that assume a simple power-law out to 100~keV (e.g., we
used the line emission file $mytl\_V000010nEp000H100\_v00.fits$).

We constructed a ``decoupled'' \texttt{mytorus} model (see, e.g.,
Yaqoob et al.\ 2012), wherein line-of-sight effects (defined to have
an inclination of $\theta = 90^{\circ}$) and reflection effects
(defined to have $\theta = 0^{\circ}$) are separated and independently
constrained.  In effect, the $\theta=90^{\circ}$ components account
for phenomena on the near side of the torus (or, obscuring geometry),
through which our line of sight passes, and the $\theta=0^{\circ}$
components account for irradiation of the face of the ``torus'' on the
far side of the central engine.

This model can be written as follows:\\

\noindent \texttt{phabs[1]*(${\rm MYTZ}_{90}$[2]*photoion[3]*
  zpowerlaw[4] $+$ zphabs[5]*const[6]* zpowerlaw[7] $+$
  const[8]*(${\rm MYTS}_{90}$[9] $+$ ${\rm MYTL}_{90}$[10]) $+$
  const[11]*(${\rm MYTS}_{0}$[12] $+$ ${\rm MYTL}_{0}$[13]) +
  mekal[14] + mekal[15] + mekal[16]})\\

\noindent Here, the components follow logically from those in the
\texttt{pexmon} model construction.  The \texttt{${\rm MYTZ}_{90}[2]$}
component is the neutral absorption component, with parameters
including the column density, inclination angle, and red-shift of the
source.  The \texttt{MYTS} and \texttt{MYTL} components are the
scattered emission and line emission components.  For these
components, the column density parameters are linked to the same
parameter in \texttt{${\rm MYTZ}_{90}[2]$}, the inclination is set to
the value indicated, and the power-law index and normalization are
linked to the same parameters in \texttt{zpowerlaw[4]}.  All of the
components above have abundances fixed at solar values.  The
photoionized absorption model is exactly the same as that employed in
the prior model, and is again required at the 4$\sigma$ level of
confidence via a simple F-test.

As with the prior \texttt{pexmon} model construction, this
\texttt{mytorus} model yields a good description of the data, and
physically reasonable parameter values (see Table 1, and Figures 4 and
5).  It is nominally a better description of the data than the
\texttt{pexmon} model, but the difference is not highly significant.
The fact that an equally good or superior fit is possible with solar
abundances suggests that super-solar abundances are not required.  In
keeping with the \texttt{pexmon} model results, scattered emission
from a dusty polar region is found to be small.

In the best-fit model, the normalization \texttt{const[11]} modulating
the \texttt{${\rm MYTS}_{90}$} and \texttt{${\rm MYTL}_{90}$}
components has a value of $0.62^{+0.07}_{-0.10}$, while that
modulating the same components viewed face-on at $\theta=0^{\circ}$
has a very small value, $4.0^{+0.3}_{-0.4}\times 10^{-3}$.  The
face-on components thus make a small contribution, but the constant is
well determined, and they are important to the fit.  Nominally, this
disparity would indicate that the reflection-like spectrum observed in NGC 4388
is dominated by scattering within the near side of the absorber
(relative to the central engine), not reflection from the far side.
However, additional fits indicate that this is partly an artificial
result, likely owing to a degeneracy that the current data cannot
break: requiring these constants to be equal only results in a
slightly worse fit ($\Delta \chi^{2} = 13$).  Given that the models
strongly require Compton-thin obscuration, making it likely that the
far side of the torus is partly visible, a conservative interpretation
of our results is that the nominal ``reflection spectrum'' is partly
reflection in the traditional sense, and partly scattering within the
near side of the absorber.  A data set that sampled a larger range in
column density, for which hardness may be a proxy, might be able to
break this degeneracy.

Fits with \texttt{mytorus} measure the same flux as the fits made with
\texttt{pexmon}, in the 0.6--10.0~keV band.  However, the best-fit
\texttt{mytorus} model implies an unabsorbed flux that is slightly
higher: $F_{\rm unabs.} = 2.63(2)\times 10^{-10}~ {\rm erg}~ {\rm
  cm}^{-2}~ {\rm s}^{-1}$.  The corresponding X-ray luminosity is
$L_{X} = 1.03(1)\times 10^{43}~ {\rm erg}~ {\rm s}^{-1}$, implying
$L_{X}/L_{\rm Edd.} = 9.7(7)\times 10^{-3}$.  The upper limit of the
range of bolometric corrections determined for Seyferts by Vasudevan
\& Fabian (2006), 70, would imply that the central engine in NGC 4388
is still just slightly sub-Eddington: $L_{\rm bol}/L_{\rm Edd.} \simeq 0.68$.

It is worth noting that our model construction applies the ionized
absorption column to flux from the central engine, but not the
scattered or line components within \texttt{mytorus}.  This
effectively assumes that the ionized gas - perhaps a wind with an
outflow velocity that is low (at least in projection) - has the bulk
of its column density interior to the neutral gas.  However, if the
ionized and neutral gas were cospatial, then the scattered and line
flux would pass through the ionized gas.  The low column, optically
thin ionized gas that we have observed would add negligible width to
the Fe K$\alpha$ line, rendering this inconsistency negligible as
well.  For much higher ionized columns, however, the inconsistency may
not be negligible.

\subsection{The Inner Extent of the ``Torus''}
Above, fits to the Fe~K$\alpha$ line with a simple Gaussian measured a
width consistent with the optical BLR in most AGN.  Indeed, studies of
narrow Fe K$\alpha$ lines observed with the {\it Chandra}/HETGS find
that approximately half have widths consistent with an origin in the
optical BLR (Shu et al.\ 2010).  In {\it Chandra} spectra of NGC 4151,
weakly relativistic dynamical imprints on the Fe~K$\alpha$ line and
its variability appear to directly constrain the line to originate in
regions even closer to the black hole than the optical BLR (Miller et
al.\ 2018, Zoghbi et al.\ 2019 in prep).

Direct measurements of the broadening expected from Keplerian orbital
motion in the BLR is likely beyond the reach of {\it NICER}.  However,
we can investigate the greatest degree of broadening that is
consistent with the {\it NICER} data, and interpret this as a lower
limit on the innermost extent of the obscuring material.  To this end,
we added an \texttt{rdblur} relativistic blurring function to modify
the \texttt{pexmon} model.  This function is analytic, and based on
the \texttt{diskline} model (e.g. Fabian et al.\ 1989), which
describes the line profile expected due to gravitational red-shifts
and strong Doppler-shifts around a Schwarzschild black hole.  This
model is not appropriate close to a spinning black hole, but
sufficient at the distances of interest in this work, and has the
advantage of extending to arbitrarily large radii.  The variable
parameters of the \texttt{rdblur} model include the line emissivity, the
inner and outer line production radii, and the inclination at which
the geometry is observed.

In our fits, the emissivity index was fixed at a value of $q=3$ (where
the emissivity is given by $J\propto r^{-q}$), and the outermost line
production radius was fixed at $r = 10^{6}~GM/c^{2}$.  We considered a
set of fixed inclination values in order to be unbiased concerning the
reflecting geometry; thus, the only variable parameter was the inner
radius.  The fit made assuming $\theta=5^{\circ}$ is consistent with
that detailed in Table 1, but the fits made assuming higher
inclination values are progressively worse.  Table 2 details the
inclinations, fit statistics, and lower radius limits obtained
throught this fitting experiment.

\section{Discussion and Conclusions}
We have analyzed a summed, time-averaged {\em NICER} spectrum of the
Seyfert-2 AGN NGC 4388, with a total exposure of 105.6~ks.  The data
have excellent sensitivity across the 0.6-10.0~keV band, permitting
strong constraints on the column density and covering factor of
neutral obscuration along our line of sight, enabling a limit on the
innermost extent of this obscuration, and facilitating the detection
of ionized absorption like that found in many Seyfert-1 AGN.  In this
section, we compare our results to prior studies of NGC 4388, and
suggest methods for further improving our knowledge of this source
with future observatories.

The values that we have measured for the equivalent neutral hydrogen
column density within the nucleus of NGC 4388 (see Table 1) are in
broad agreement with measurements made using {\it Suzaku} and {\it
  XMM-Newton}.  The {\it Suzaku} data were analyzed by Shirai et
al.\ (2008); using two related models, values of $N_{\rm H} =
2.36(6)\times 10^{23}~ {\rm cm}^{-2}$ and $N_{\rm H} =
2.51^{+0.04}_{-0.03}\times 10^{23}~ {\rm cm}^{-2}$ were measured.
Fits to the spectra obtained in two {\it XMM-Newton}/EPIC observations
of NGC 4388 measured column densities of $N_{\rm H} =
2.45^{+0.20}_{-0.21}\times 10^{23}~ {\rm cm}^{-2}$ and $N_{\rm H} =
2.79(7)\times 10^{23}~ {\rm cm}^{-2}$ (Beckmann et al.\ 2004).

A {\it NuSTAR} observation of NGC 4388 made in late 2013 was analyzed
by Kamraj et al.\ (2017).  Two measurements of the absorbing column
were obtained in separate fits to the {\it NuSTAR} data: $N_{\rm H} =
6.5(8)\times 10^{23}~ {\rm cm}^{-2}$, and $N_{\rm H} = 5.3(7)\times
10^{23}~ {\rm cm}^{-2}$; the former was made using \texttt{pexrav},
and the latter using \texttt{mytorus}.  As noted above,
\texttt{pexrav} does not include line emission, though the strength of
the Fe~K$\alpha$ line relative to the edge is set by atomic physics
and this scaling is key to accurately constraining the crucial
parameters.  The \texttt{mytorus} modeling undertaken by Kamraj et
al.\ (2017) included only a single scattering and line component
group, and a single inclination (fixed to the viewing angle of the
accretion flow).  

Masini et al.\ (2016) had previously analyzed the {\it NuSTAR} spectra
of NGC 4388 as part of a larger analysis of water maser AGN.  They
examined a small set of self-consistent \texttt{mytorus} models,
including one that is essentially the same as our implementation.
That treatment resulted in an obscuring column of $N_{\rm H} = 4.2(5)
\times 10^{23}~{\rm cm}^{-2}$.  The fact that a higher column is
measured using the same model suggests that the difference could
reflect a genuine reduction in the obscuring column between the {\it
  NuSTAR} observations in 2013, and our more recent {\it NICER}
program.  It is interesting to note that -- similar to our findings --
Masini et al.\ (2016) also report that scattering within the obscuring
gas between the central engine and the observer dominates over
forward-scattering from the same material on the far side of the
central engine.

In the absence of a direct density constraint, writing $N_{\rm H} =
nr$ allows an upper limit on the absorption radius via the ionization
parameter, $\xi = L/Nr$.  As noted above, and in Table 1,
$L_{x}/L_{\rm Edd}$ is  0.0078--0.0104.  A modest
bolometric correction is likely more appropriate, based on the work of
Vasudevan \& Fabian (2007).  Adopting a correction factor of 20 then
implies $L_{\rm bol} \simeq 2\times 10^{44}~ {\rm erg}~ {\rm s}^{-1}$.
Using this value, and the absorption parameters listed in Table 1, the
ionization parameter formalism implies a radius limit of $r\leq
5\times 10^{18}~{\rm cm}$, or $r\leq 4\times 10^{6}~GM/c^{2}$ for
the mass of NGC 4388.  However, assuming a density of $n=10^{8}~{\rm
  cm}^{-3}$ -- consistent with ionized absorber density constraints in
some Seyferts (e.g., Krongold et al.\ 2007) -- a radius of $r \simeq
2\times 10^{4}~GM/c^{2}$ results, broadly consistent with an origin
in the optical BLR.  There, it might help to pressure-confine cold,
clumpy gas.  We note that the photoionization model assumed a
power-law index of $\Gamma = 1.7$ whereas values consistent with
$\Gamma = 1.55$ are measured, but this difference is minor compared to
uncertainties in the bolometric correction factor.

Using the neutral reflected spectrum as a trace of the innermost
extent of the cold obscuration, we are able to estimate lower limits
on the innermost extent of this absorption.  The smallest value,
$r\geq 270~GM/c^{2}$, is 1--2 orders of magnitude smaller than a typical
optical BLR.  If the reflector in NGC 4388 originates at such small
radii, it would likely trace an X-ray BLR, warp, or other geometry.  A
larger limit that is not strongly rejected by an increased fit
statistic, $r\geq 1600~GM/c^{2}$ (corresponding to the fit made
assuming $\theta = 30^{\circ}$), is consistent with the inner extent
of the optical BLR in many sources.

Especially in galaxies for which our line of sight passes close to the
plane of the galactic disk, such as NGC 4388, the host galaxy itself
can contribute significantly to the gaseous and dust obscuration
detected in Seyferts and even some Compton-thick AGN (see, e.g., Shi
et al.\ 2006, Gandhi et al.\ 2009, Goulding et al.\ 2012, Puccetti et
al.\ 2014).  It is therefore possible that the host galaxy contributes
to the gaseous column density observed in NGC 4388, but the bulk of
the neutral and ionized column likely arises on much smaller scales,
based on the variability, line widths, and ionization arguments
discussed above.  However, Shi et al.\ (2006) measure a strong
9.7~$\mu$m silicate feature in NGC 4388, and {\em Hubble} images of
NGC 4388 reveal strong spiral arms and prominent dust lanes (Greene et
al.\ 2013); the host galaxy could contribute significantly to the dust
obscuration.

If the majority of the gaseous obscuration occurs on small scales within
the nucleus, and if at least some part of the dust obscuration occurs
within the host galaxy, is there a role for a conventional
parsec-scale torus in NGC 4388?  Dust reverberation mapping --
actually reverberation using the K-band continuum at 2.2$\mu$m because
it is suitably close to the peak emissivity of hot dust grains --
finds clear evidence of dust on parsec scales (e.g., Koshida et al.\
2014) in nearby Seyferts.  However, mid-IR interferometry has shown
that the bulk of the IR emission occurs in polar regions rather than
in equatorial regions (e.g., Raban et al.\ 2009; Honig et al.\ 2012,
2013).  This may indicate that in NGC 4388 and other Seyferts, a
traditional torus geometry is only a minor factor in shaping the
broadband SED; however, this does not mean that different obscuration
zones are completely disconnected.  Recent studies of several AGN link
X-ray activity in the central engine to dust flows on galactic scales
(e.g., Cicone et al.\ 2014).

Future missions will combine spectral resolution and sensitivity, and
may be able to build on the results achieved with {\it NICER}.  Figure
6 shows simulated {\it XRISM/Resolve} (Tashiro et al.\ 2018) and {\it
  Athena/X-IFU} (Barret et al.\ 2018) spectra of NGC 4388, based
on the results presented in this work.  The {\it XRISM/Resolve}
spectra were simulated using {\it Hitomi} responses with a resolution
of 5~eV.  Based on the best-fit \texttt{mytorus} model in Table 1, if
the reflector traces the innermost extent of the obscuration and
originates at $r = 1600~GM/c^{2}$ (see above), dynamical broadening
would be detected in a 100~ks exposure.  The X-IFU simulation was
constructed using the current public response matrices, with a
resolution of 2~eV.  We find that the {\em X-IFU} will be able to
detect small changes (factors of $\sim2$) in the high resolution
features tied to the neutral and ionized absorbers on time scales as
short as 10~ks, commensurate with the variability reported in Elvis et
al.\ (2004) but based on low-resolution data.  This would represent a
very strong confirmation of the prior variability, and open a window
into the nature of obscuration in Seyferts on dynamical time scales.

Finally, it is worth noting some of the limitations of our spectral
modeling.  Both \texttt{pexmon} and \texttt{mytorus} assume cold,
neutral gas.  If the gas that drives obscuration along our line of
sight (and, reflection from the far side of the central engine) is as
close as the optical BLR, the gas may not be entirely neutral.  A
model such as \texttt{xillver} (Garcia et al.\ 2013) describes
reflection from a broad range of gas ionization levels.  Exploratory
fits with \texttt{xillver} replacing \texttt{pexmon}, for instance,
yield equally good fits.  Whereas \texttt{pexmon} only includes lines
from Fe and Ni, \texttt{xillver} includes a broad range of abundant
elements, and contributes some line flux (particularly Si) that is
described by the \texttt{mekal} components in our models.  However, more
self-consistent implementations of \texttt{xillver} that replace the
neutral obscuration with ionized obscuration through \texttt{zxipcf},
yield significantly worse fits.  This may indicate that the absorption
and reflection at least occur in a mixed medium, potentially with an
ionization gradient.  Future refinements in the calibration of {\em
NICER}, and future missions, may be able to address such points.

We thank the anonymous referee for comments that improved the paper.
JMM acknowledges helpful conversations with Richard Mushotzky.

\clearpage


\begin{figure}
\hspace{-0.2in}
\includegraphics[scale=0.8]{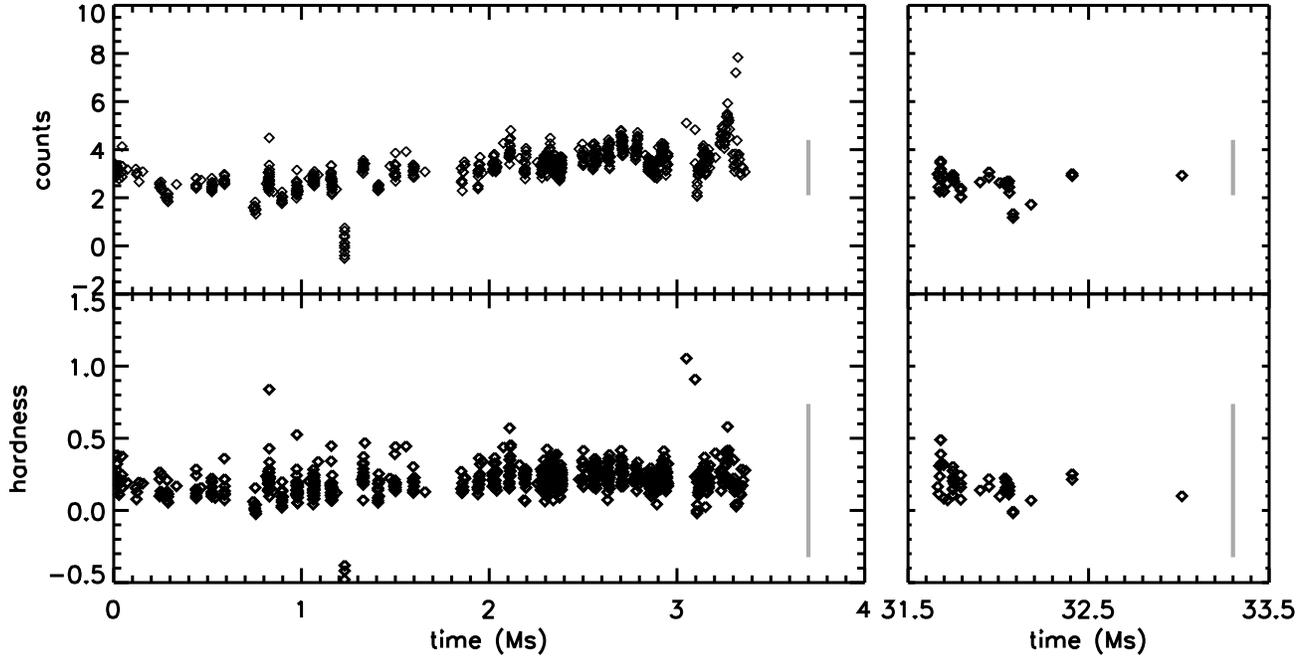}
\figcaption[t]{\footnotesize The {\em NICER} light curve and hardness
  ratio of NGC 4388, following our initial screening.  The time axis
  is split to better represent two intervals of dense monitoring that
  were separated by a long period without monitoring.  In the counts
  versus time panels, each point represents the background-subtracted
  number of counts in the 0.3--12.0~keV band per detector in 128~s of
  integration.  The hardness curve is the ratio of the counts in the
  8--12 keV band, to the counts in the 0.3--2.0 keV band.  The thick
  gray line in each panel represents the median error, plotted at the
  median count rate (or, hardness).  The low points at 1.2~Ms are due
  to a background flare; these intervals were excised.  The remaining
  extremal flux points in the light curve are rare and confined to
  relatively few 128~s intervals.  The errors on the hardness curve
  are necessarily large but the central values are fairly steady
  across the monitoring period.}
\end{figure}
\medskip

\clearpage

\begin{figure}
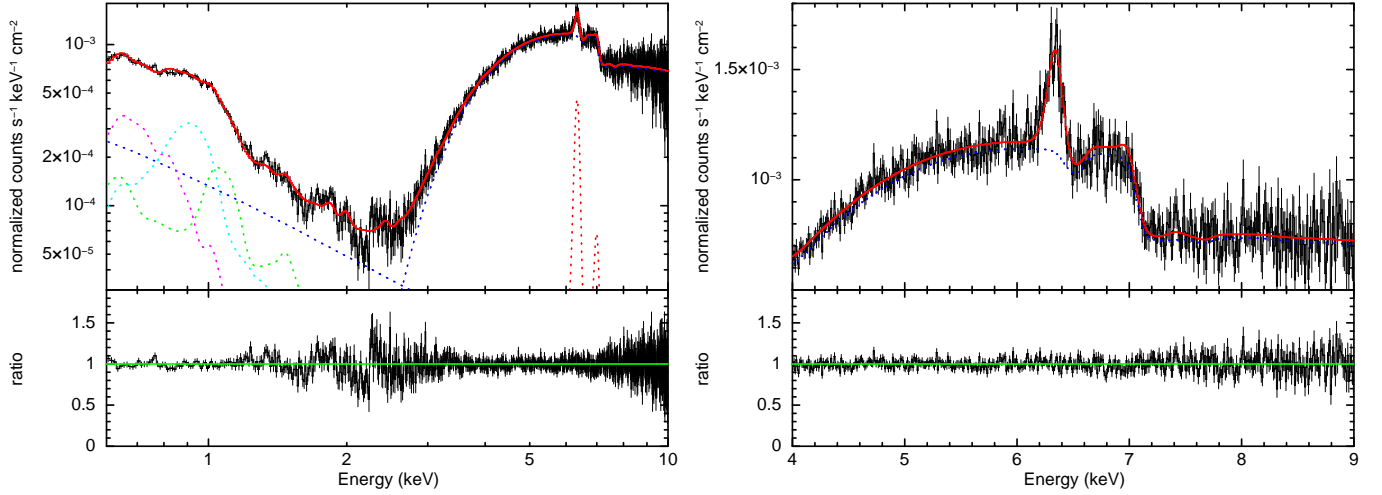

\hspace{-0.2in}
\includegraphics[scale=0.35,angle=-90]{f2a.ps}
\hspace{-0.2in}
\includegraphics[scale=0.35,angle=-90]{f2b.ps}
\figcaption[t]{\footnotesize The {\em NICER} spectrum of NGC 4388.
  The data are fit with a model including Compton-thin neutral
  absorption and ionized absorption along the line of sight, distant
  neutral reflection via \texttt{pexmon}, and diffuse plasma from the
  extended nuclear region at low energy (see Table 1).  The data were
  binned to have a signal-to-noise ratio of 10.0 prior to fitting, and
  they are shown in the rest frame of NGC 4388 ($z=0.085$).  The total
  best-fit model is shown in solid red, the direct power-law is shown
  in dashed blue above 3 keV, the scattered power-law is shown in
  dashed blue below 3 keV, reflection is shown in dashed red, and the
  three plasma components are shown in dashed magenta, cyan, and green below
  2 keV.  {\bf LEFT}: The {\it NICER} spectrum and data/model ratio
  shown on the full 0.6--10.0~keV fitting band.  {\bf RIGHT}: The {\it
    NICER} spectrum and data/model ratio in the Fe~K band.  Note the
  very strong neutral Fe K$\alpha$ emission line at 6.40~keV and
  associated edge at 7.05~keV.  Weak, ionized absorption in the
  6.6--6.7~keV range was modeled with XSTAR and is required at the
  $4\sigma$ level of confidence.}
\end{figure}
\medskip

\begin{figure}
\includegraphics[scale=0.6,angle=-90]{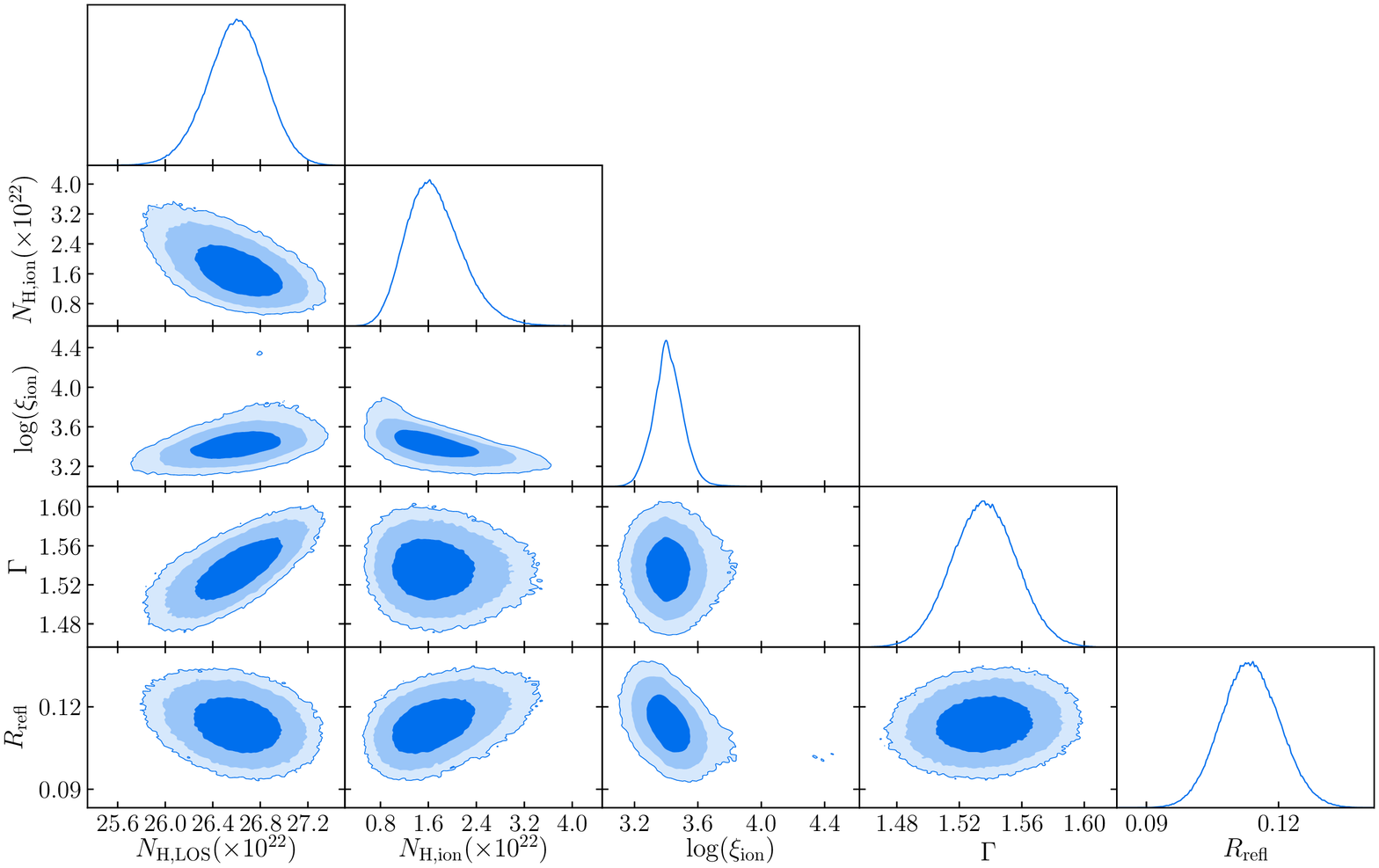}
\vspace{-0.3in}
\figcaption[t]{\footnotesize Confidence contours from the MCMC
  analysis related to the fits with \texttt{pexmon} (see Table 1 and
  the text).  The progressively lighter blue hues indicate the
  $1\sigma$, $2\sigma$, and $3\sigma$ levels of confidence.  The
  final panel in each row depicts the normalized one-dimensional
  probability density function for the parameter of interest.}
\end{figure}
\medskip

\clearpage

\begin{figure}
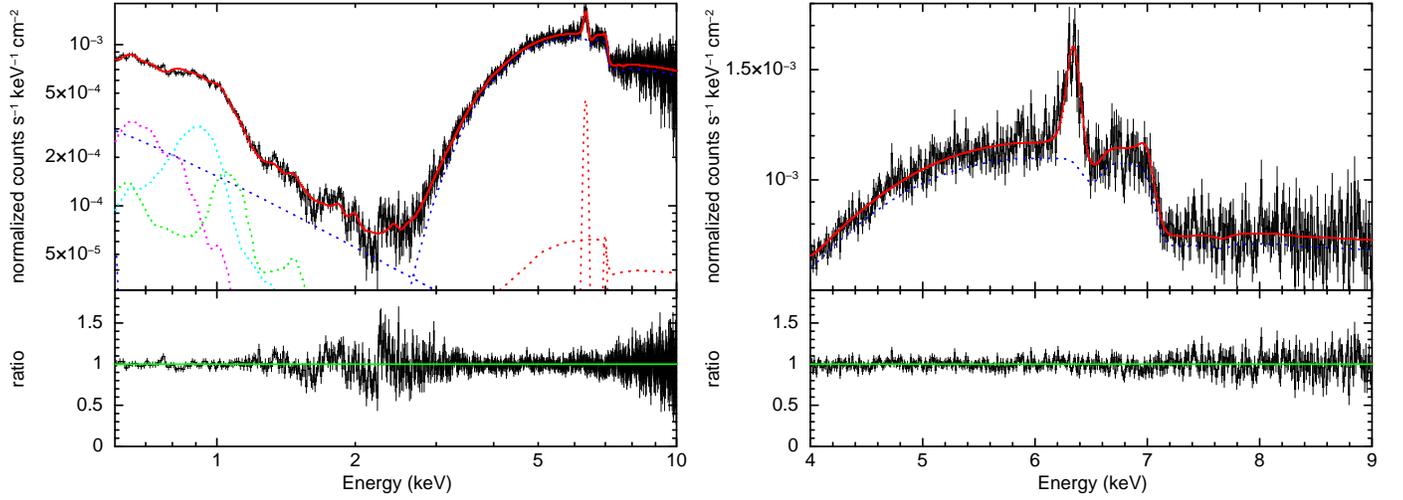

\hspace{-0.2in}
\includegraphics[scale=0.35,angle=-90]{f4a.ps}
\hspace{-0.2in}
\includegraphics[scale=0.35,angle=-90]{f4b.ps}
\figcaption[t]{\footnotesize The {\it NICER} spectrum of NGC 4388,
  shown as per Figure 2, but independent neutral absorption and
  distant reflection (\texttt{pexmon}) have been replaced with the
  \texttt{mytorus} suite, including self-consistent neutral
  absorption, continuum scattering, and lines from distant irradiated
  material (see Table 1).  The total best-fit model is shown in solid
  red, the direct power-law is shown in dashed blue above 3 keV, the
  scattered power-law is shown in dashed blue below 3 keV, reflection
  is shown in dashed red, and the three plasma components are shown in
  dashed magenta, cyan, and green below 2 keV.  {\bf LEFT}: The {\it NICER}
  spectrum and data/model ratio shown on the full 0.6--10.0~keV
  fitting band.  }
\end{figure}
\medskip

\begin{figure}
\includegraphics[scale=0.6,angle=-90]{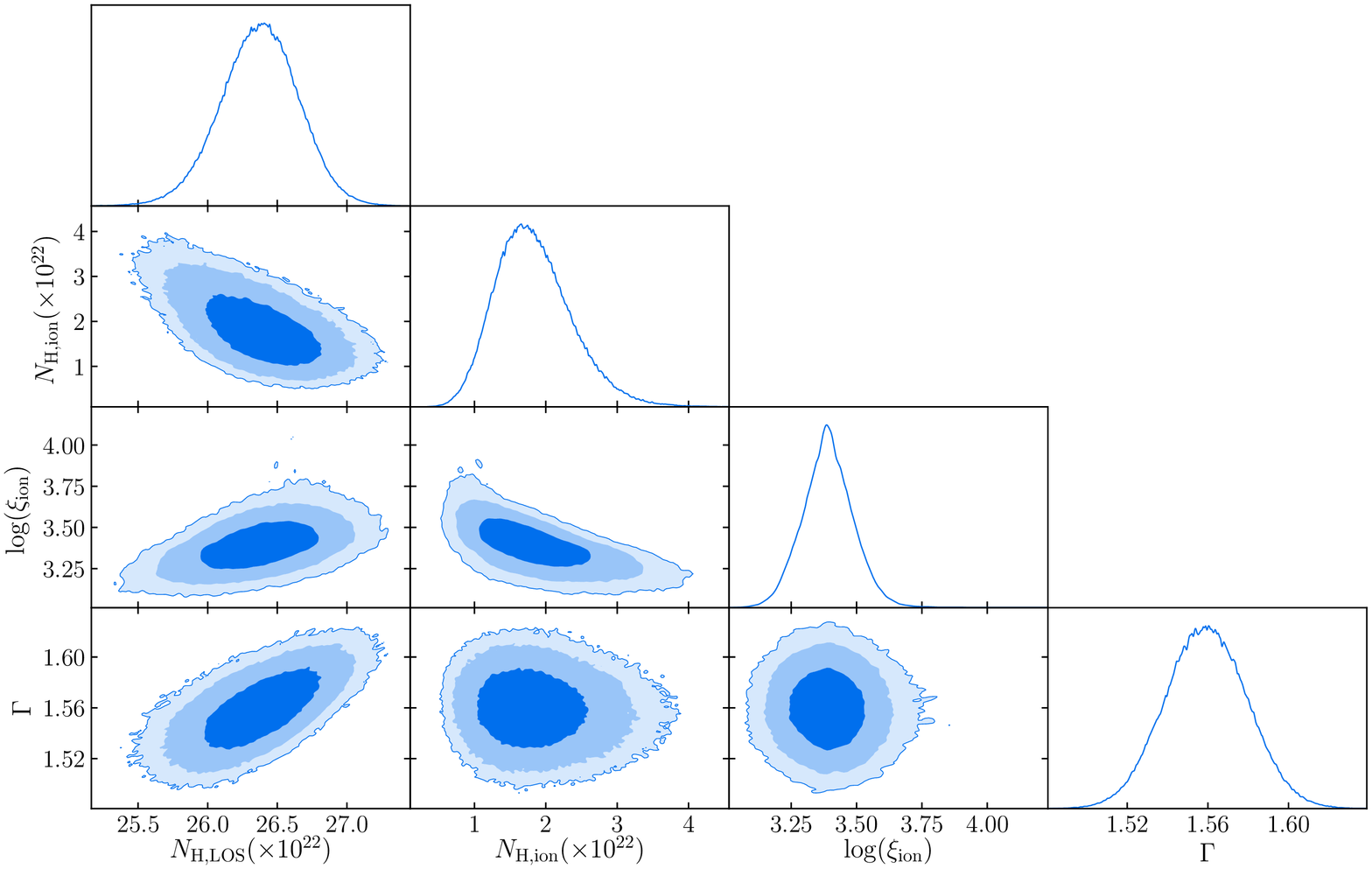}
\vspace{-0.3in}
\figcaption[t]{\footnotesize Confidence contours from the MCMC
  analysis related to the fits with \texttt{mytorus} (see Table 1 and
  the text).  The progressively lighter blue hues indicate the
  $1\sigma$, $2\sigma$, and $3\sigma$ levels of confidence.  The
  final panel in each row depicts the normalized one-dimensional
  probability density function for the parameter of interest.}
\end{figure}
\medskip

\clearpage

\begin{table}[t!]
\caption{Spectral Fitting Results}
\begin{center}
\begin{tabular}{lll}
Parameter                                &      \texttt{pexmon}     &      \texttt{mytorus} \\
\hline
$N_{\rm H, LOS} (10^{23}~{\rm cm}^{-2})$  &  $2.67_{-0.03}^{+0.02}$  &              $2.64(3)$ \\

$N_{\rm H, LOS, ion} (10^{22}~{\rm cm}^{-2})$ &    $1.7(5)$           &           $1.8_{-0.5}^{+0.6}$  \\
${\rm log} \xi$                              &     $3.44_{-0.12}^{+0.07}$  &       $3.4(1)$ \\

$\Gamma$                                 &  $1.54(2)$  &        $1.56(2)$ \\
$K_{PL}~ (10^{-2})$                         &         3.2(1)         &       $3.96(2)$  \\

$\theta_{\rm inc}$                          &     $5^{\circ}$            &     $90^{\circ}$       \\       
$\theta_{\rm torus}$                       &      $5^{\circ}$          &      $0^{\circ}$         \\

$R_{\rm refl}~ (10^{-1})$                  &    $1.13_{-0.07}^{+0.06}$    &          --            \\

$f_{SC+lines}~ (10^{-1})$                &            --           &         $6.2_{-1.0}^{+0.7}$ \\

$N_{\rm H, SC, poles}~ (10^{20}~{\rm cm}^{-2}) $ &     --            &        $\leq 2.0$  \\  
$f_{SC, poles}~ (10^{-3})$                 &    $3.5^{+0.2}_{-0.7}$   &     $4.0^{+0.3}_{-0.4}$ \\

$kT_{\rm mekal, 1}$~ (keV)                 &         0.29(1)         &         $0.31(1)$ \\
$K_{\rm mekal, 1}~ (10^{-5})$              &   $8.5^{+0.3}_{-0.5}$   &         $8.2(5)$ \\

$kT_{\rm mekal, 2}$~ (keV)                 &   $0.81_{-0.04}^{+0.02}$    &         $0.82(3)$     \\
$K_{\rm mekal, 2}~ (10^{-5})$              &   $9.5^{+0.6}_{-0.5}$   &           $9.3_{-0.6}^{+0.4}$ \\

$kT_{\rm mekal, 3}$~ (keV)                 &   $1.8^{+0.4}_{-0.1}$   &         $1.9_{-0.2}^{+0.3}$ \\
$K_{\rm mekal, 3}~ (10^{-5})$              &   $20^{+6}_{-2}$        &         $21^{+5}_{-3}$ \\

\hline

F$_{0.6-10}$~$(10^{-11}~{\rm erg}~{\rm cm}^{2}~{\rm s}^{-1})$ &  6.2(2)          &      6.23(4)           \\
F$_{0.6-10, unabs.}$~$(10^{-10}~{\rm erg}~{\rm cm}^{2}~{\rm s}^{-1})$  &  2.21(7)         &     2.63(2)       \\
L$_{{\rm X}, 0.6-10}$~$(10^{43}~{\rm erg}~{\rm s}^{-1})$                &  0.86(3)         &   1.03(1)           \\
L$_{\rm X}$/L$_{Edd.}$~$(10^{-3})$                                    &   8.1(3)          &    9.7(7)       \\

\hline

$\chi^{2}/\nu$                          &   $1123.6.0/926$            &         $1116.4/924$     \\
p$_{\rm KS}$                                &    60.6\%                  &           65.0\%          \\

\hline

\hline
\end{tabular}
\vspace{-0.1in}
\vspace*{\baselineskip}~\\ \end{center} 
\tablecomments{Key spectral model parameters and $1\sigma$ confidence
  errors, derived from fits to the 0.6--10~keV {\it NICER} spectrum of
  NGC 4388.  The models consist of neutral and ionized obscuration
  within the nucleus ($N_{\rm H, LOS}$ and $N_{\rm H, LOS, ion}$),
  covering intrinsic power-law emission (with a photon index $\Gamma$,
  and flux normalization $K_{PL}$).  The exact details of the
  obscuration, transmission, and reflection of the emission from the
  central engine depend on the nuances of the \texttt{pexmon} and
  \texttt{mytorus} models and their implementation; this is especially
  true of the angles related to the assumed geometry ($\theta_{\rm
    inc}$ and $\theta_{\rm torus}$).  Please see the text for full
  details.  The \texttt{pexmon} model includes a reflection fraction
  (relative to the direct continuum, $R_{\rm refl}$); this flux is
  separately captured in scattered and line model components by
  \texttt{mytorus} and the value of the normaliztion for these
  components is given ($f_{SC+lines}$).  Both of our model
  constructions also allow for flux to be scattered by a polar region
  with a distinct column density (described by the parameters $N_{\rm
    H, SC, poles}$ and $f_{SC, poles}$).  Finally, three
  \texttt{mekal} plasma components are used to describe diffuse
  emission in the nucleus and surrounding galaxy (each has a unique
  temperature and normalization).  Where parameters lack errors, the
  parameter was fixed in the fit.  The measured flux, unabsorbed flux,
  and X-ray luminosity as a fraction of the Eddington luminosity
  (assuming $M = 8.4\pm 0.2 \times 10^{6}~M_{\odot}$, Kuo et
  al.\ 2011) are also reported.  The bolometric correction to the
  X-ray luminosity could be as low as 15 and as high as 70 (see, e.g.,
  Vasudevan \& Fabian 2007).  Errors are 1$\sigma$ confidence errors,
  based on a large MCMC exploration of the parameter space.  Symmetric
  errors are indicated in parentheses and reflect the uncertainty in
  the last significant digit.  The final rows give the $\chi^{2}$
  divided by the degrees of freedom $\nu$ from a simple fit with these
  model parameters.  The goodness of the fits in the MCMC analysis was
  evaluated using the Kolmogorov-Smirnov (KS) test in XSPEC.  The
  $p_{\rm KS}$ parameter represents the percentage of 100 simulations
  that produced a KS statistic less than the value for the best-fit
  model.}
\vspace{-0.15in}
\end{table}
\smallskip

\begin{table}[t!]
\caption{Lower limits on the Fe~K line production region}
\begin{center}
\begin{tabular}{lll}
Inclination  &  radius limit    &  $\chi^{2}$ \\
(deg.)       &  ($GM/c^{2}$)    &     ~     \\
\hline
$5^{\circ}$   &  $r > 270$       &    1123.5 \\
$30^{\circ}$  &  $r > 1600$      &    1126.2 \\
$45^{\circ}$  &  $r > 2500$      &    1128.2 \\
$60^{\circ}$  &  $r > 4500$      &    1133.3 \\
\hline
\end{tabular}
\vspace{-0.1in}
\vspace*{\baselineskip}~\\ \end{center} 
\tablecomments{Lower limits on the Fe~K line production region.  The
  limits were obtained by convolving the \texttt{pexmon}-based model
  witht the ``rdblur'' function, considering a small set of plausible
  inclinations at which the neutral reflector (obscuring gas) might be
  viewed in the far side of the central engine.  The limits given are
  90\% confidence lower limits, and the value of the $\chi^{2}$ fit
  statistic on that boundary is also listed.}
\vspace{-0.15in}
\end{table}
\smallskip

\clearpage

\begin{figure}
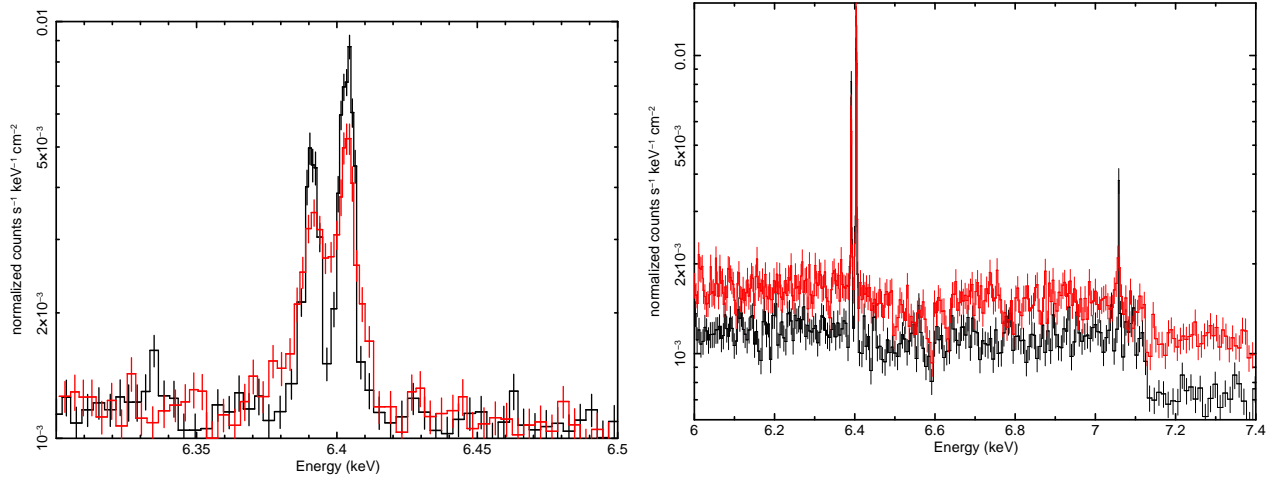

\begin{center}
\hspace{-0.2in}
\includegraphics[scale=0.35,angle=-90]{f6a.ps}
\hspace{-0.2in}
\includegraphics[scale=0.35,angle=-90]{f6b.ps}
\end{center}
\figcaption[t]{\footnotesize {\bf LEFT}: A simulated 100~ks {\it
    XRISM}/Resolve spectrum of NGC 4388, based on the best-fit
  \texttt{mytorus} model shown in Figure 4 and detailed in Table 1.
  The simulated spectrum was binned to require a signal-to-noise ratio
  of 10.  Note that the Fe~K$\alpha$ line is revealed as a doublet.
  The spectrum in black is consistent with a distant, parsec-scale
  torus, whereas the spectrum in red (with broader lines) assumes
  emission from $r = 1600~GM/c^{2}$, a radius allowed by the {\it
    NICER} spectrum and consistent with prior variability (Elvis et
  al.\ 2004).  {\bf RIGHT}: Simulated 10~ks {\it Athena}/X-IFU spectra
  of NGC 4388.  The spectrum in black was generated assuming the same
  model depicted in Figure 4 and detailed in Table 1.  The spectrum in
  red has half of the neutral absorption column density and twice the
  ionized absoprtion column density.  The weaker Fe~K edge and
  stronger absorption lines in the red spectrum are readily apparent
  and suggest that future observations of the Fe~K band alone may be
  sufficient to measure column densities.}
\end{figure}
\medskip

\end{document}